\documentclass[times,authoryear]{elsarticle}

\usepackage[utf8]{inputenc}
\usepackage{graphicx}
\usepackage{url}
\graphicspath{ {images/} }
\usepackage{caption}
\usepackage{amsmath}
\usepackage{verbatim}

\usepackage{dingbat} 

\usepackage{subfigure}

\usepackage{orcidlink}

\usepackage[authoryear]{natbib}

\usepackage{siunitx} 

\RequirePackage{fix-cm}

\usepackage{tablefootnote}
\usepackage{pdflscape}

\def\tsc#1{\csdef{#1}{\textsc{\lowercase{#1}}\xspace}}
\tsc{WGM}
\tsc{QE}
\tsc{EP}
\tsc{PMS}
\tsc{BEC}
\tsc{DE}

\usepackage{hyperref}
\hypersetup{
    colorlinks=true,
    linkcolor=blue,
    filecolor=magenta,      
    urlcolor=cyan,
    pdftitle={Overleaf Example},
    pdfpagemode=FullScreen,
    }    
\usepackage{booktabs}
\begin{document}

\title{Solarsystem: A Validated Lightweight Python Package for Planetary Positions and Solar-Lunar Event Calculations}

\author{Ioannis Nasios}


\begin{frontmatter}

\begin{abstract}
This paper presents \texttt{solarsystem}, a validated lightweight and dependency-free Python package for planetary positions and solar-lunar event calculations. The package provides heliocentric and geocentric positions for the major planets, selected dwarf planets, the Centaur Chiron, and the Moon, together with sunrise, sunset, moonrise, moonset, and lunar illumination calculations. Additional functionality includes coordinate transformations between commonly used astronomical reference systems.

The implemented algorithms employ analytical models that avoid reliance on external ephemeris datasets, resulting in a portable and computationally efficient solution suitable for a broad range of astronomical applications. An optional precession correction model is included, enabling calculations either in a precession-corrected reference frame or in a fixed epoch framework, depending on user requirements.

The numerical performance of \texttt{solarsystem} was evaluated against the JPL DE440 planetary ephemerides using the \texttt{Skyfield} framework as a reference. Validation experiments spanning multiple bodies and extended temporal intervals demonstrate good agreement with the reference ephemerides, with mean planetary longitude and latitude deviations of approximately 0.44 and 0.16 arcminutes, respectively. Additional validation of solar and lunar event calculations yielded timing differences of only a few minutes relative to the reference solutions, while lunar illumination estimates differed by approximately 0.2\%.

The package can be installed directly through PyPI while the source code, documentation, validation notebooks and example workflows are publicly available through the project repository in \url{https://github.com/IoannisNasios/solarsystem}.
\end{abstract}

\end{frontmatter}

\section{Introduction}
\label{sec:introduction}
At its core, solarsystem provides a lightweight framework for heliocentric (Sun-centered) and geocentric (Earth-centered) astronomical calculations involving a broad collection of Solar System bodies. The library supports computations for all eight major planets, ranging from Mercury to Neptune, in addition to several recognized dwarf planets such as Pluto, Ceres and Eris. Selected minor bodies as Centaur Chiron, are also incorporated into the computational framework. Furthermore, the package includes dedicated routines for calculating the apparent position and phase-related properties of Earth's natural satellite, the Moon, as observed from a terrestrial reference location.

The project is distributed under the MIT License and is publicly available through its GitHub repository, with installation provided directly through the Python Package Index (PyPI). While many contemporary astronomical software packages achieve high precision through the use of external ephemerides, numerical kernels, or extensive scientific software dependencies, such approaches can increase installation, deployment, and computational complexity. In contrast, \texttt{solarsystem} adopts a lightweight analytical approach that enables the computation of planetary positions and solar-lunar events without requiring external datasets. The package is therefore intended to provide a computationally efficient and portable solution for Solar System calculations while maintaining good agreement with modern reference ephemerides.

An optional precession correction field is also included within \texttt{solarsystem}. By default, precession effects associated with the gradual axial motion of the Earth \citep{berger1978long} are taken into account in positional calculations. However, users may explicitly disable this behavior through the \texttt{precession=False} option, allowing celestial coordinates to remain referenced to a fixed frame without applying epoch-dependent axial corrections. This functionality can be important for applications requiring stable long-term relative positioning or direct comparison within a constant coordinate framework. In contrast, modern ephemerides generally incorporate precession and related coordinate evolution effects by default. Providing the ability to selectively deactivate precession therefore offers additional flexibility for educational demonstrations, visualization workflows, simplified orbital studies, and comparative numerical analyses where a non-precessing reference frame may be desirable. The package has additionally been employed in peer-reviewed astronomical research, demonstrating its utility beyond educational and visualization-oriented applications \citep{kuang2023simulations}.

\subsection{Package Overview}
\label{system_overview}

\subsubsection{Celestial Body Positioning}
A principal capability of solarsystem is the computation of positional information for Solar System bodies. The package enables users to calculate heliocentric and geocentric coordinates, including longitude, latitude, and radial distance, for planets, dwarf planets, and selected small bodies at arbitrary dates and times. Furthermore, it enables geocentric coordinate calculations for the Moon.

These calculations allow the library to be used in a variety of astronomical contexts, including orbital visualization, educational demonstrations, and scientific analyses. The resulting coordinate data can also serve as input for external plotting frameworks, simulation environments, or numerical analysis pipelines.

Although the current implementation includes a limited set of dwarf planets (Pluto, Ceres, and Eris) and a single Centaur object (Chiron), the underlying computational framework is not restricted to these bodies. The adopted implementation strategy illustrates how orbital elements can be incorporated into the analytical model, providing a template for the inclusion of additional dwarf planets, Centaurs, trans-Neptunian objects, or other minor bodies. Consequently, the package serves not only as a computational tool but also as a flexible framework that can be readily extended as new requirements arise. Future releases are expected to expand the catalog of supported objects.

\subsubsection{Astronomical Event Calculation}

Beyond positional astronomy, solarsystem includes modules dedicated to the computation of common astronomical events associated with solar and lunar cycles. These include:

\begin{description}
    \item[Sunrise and sunset] times for arbitrary geographic coordinates,
    \item[Moonrise and moonset] times for arbitrary geographic coordinates,
    \item[Lunar illuminated fraction] calculations.
\end{description}

Such functionality is particularly useful for observational planning, astronomical calendar generation, and educational demonstrations involving Earth-Moon-Sun geometry. By combining positional calculations with event timing utilities, the library offers a compact toolkit for practical observational astronomy applications.

\subsubsection{Coordinate Transformations}

Astronomical computations frequently require transformations between multiple spatial reference systems. Supporting this workflow, a set of functions for converting between commonly used coordinate representations are provided. These include transformations between:

\begin{description}
    \item[Spherical and rectangular (Cartesian)] coordination systems, and
    \item[Ecliptic and equatorial] planes.
\end{description}

These tools facilitate interoperability with visualization software, external astrophysical libraries, and custom numerical workflows that rely on alternative coordinate conventions.

\subsection{Scientific and Observational Applications}
\label{Scientific and Observational Applications}

The lightweight architecture of solarsystem make it suitable for a wide range of scientific, educational, and exploratory applications.

\begin{itemize}

\item \textit{Observational astronomy}. Generation of sunrise, sunset, moonrise, moonset, and lunar illumination datasets for site-specific observing conditions.

\item \textit{Computational astronomy}. Rapid generation of heliocentric and geocentric positional data for exploratory orbital analysis, simulation studies, and comparison against reference ephemerides.

\item \textit{Visualization and monitoring}. Creation of time-dependent Solar System visualizations, planetary configuration studies, and long-term astronomical calendars.

\end{itemize}

\section{Related Work and Background}
\label{lierature}

The increasing adoption of Python within the scientific computing community has led to the development of numerous astronomy- and astrodynamics-oriented software frameworks. These tools range from high-precision ephemeris engines and orbital mechanics libraries to lightweight educational packages aimed at visualization and exploratory analysis. In recent years, the open-source ecosystem surrounding astronomical software has expanded considerably, enabling researchers, educators, and hobbyists to access computational tools that were previously restricted to specialized scientific environments.

One of the most widely adopted astronomy frameworks in Python is Astropy \citep{astropy2013}, which provides a comprehensive ecosystem for astronomical data analysis, coordinate transformations, time handling, and observational calculations. Astropy has become a foundational dependency for many astronomy-related projects due to its robust implementation of coordinate systems, units, and interoperability standards. However, its broad scope and extensive functionality can introduce additional complexity for lightweight applications or educational use cases where only a subset of astronomical computations is required.

In the domain of astrodynamics and orbital mechanics, libraries such as poliastro \cite{poliastro} provide sophisticated tools for orbit propagation, mission analysis, and trajectory visualization. These frameworks are highly valuable for aerospace engineering and scientific research but often depend on external numerical libraries and detailed physical models. Similarly, NASA's SPICE toolkit and its Python interfaces, such as SpiceyPy \citep{spiceypy}, enable highly accurate spacecraft geometry and ephemeris computations through the use of externally supplied SPICE kernels. 

Other software efforts have focused on planetary visualization and observational astronomy. Projects such as Skyfield \citep{skyfield} simplify ephemeris-based calculations using modern astronomical datasets derived from the Jet Propulsion Laboratory (JPL). These tools prioritize accurate positional astronomy while remaining relatively accessible to non-specialist users. Nevertheless, many such packages still rely on external ephemeris downloads and supporting numerical infrastructures.

The development and publication of scientific software has become an increasingly important component of modern astronomical research. Representative examples include \textit{GalSim}, a modular framework for astronomical image simulation \citep{rowe2015galsim}, and \textit{HOPE}, a Python-based framework for accelerating astrophysical computations through just-in-time compilation \citep{akeret2015hope}. These studies illustrate the growing recognition of computational frameworks as research contributions in their own right, emphasizing software architecture, performance characterization, validation, and reproducibility. The present work follows a similar philosophy by describing the design, validation, and performance characteristics of a lightweight framework for Solar System calculations and astronomical event prediction.

Within the broader ecosystem of astronomical software, \texttt{solarsystem} occupies a niche focused on lightweight analytical computation, portability, and ease of deployment. Rather than targeting professional-grade astrodynamics or sub-arcsecond astrometric precision, the framework is designed to support rapid Solar System calculations, astronomical event prediction, and exploratory numerical analyses without reliance on external ephemerides, datasets, or third-party scientific libraries. By implementing core heliocentric and geocentric calculations directly in pure Python, the package provides a computationally efficient and reproducible environment for planetary position calculations and related astronomical workflows. Such an approach remains valuable in applications where simplicity, transparency, and portability are prioritized alongside scientifically meaningful levels of accuracy.

In addition to its educational and visualization-oriented applications, \texttt{solarsystem} has already been employed in peer-reviewed astronomical research. \cite{kuang2023simulations} utilized solarsystem package in their investigation of the detectability of triple microlensing events involving a scaled Sun-Jupiter-Saturn system, published in the \textit{Monthly Notices of the Royal Astronomical Society}. The use of \texttt{solarsystem} within a peer-reviewed astrophysical investigation demonstrates that lightweight analytical frameworks can provide practical utility in scientific simulation and modeling workflows.

\section{System Design and Implementation}
\label{system_design}

\subsection{Mathematical Background}
\label{mathback}

The computational framework implemented in solarsystem is based on classical celestial mechanics and spherical astronomy \citep{meeus1991astronomical}. Planetary positions are determined through analytical approximations of orbital motion using time-dependent orbital elements referenced to a common epoch \citep{simon1994numerical}. These orbital parameters are used to estimate heliocentric positions of Solar System bodies, which can subsequently be transformed into geocentric coordinates for Earth-based observational calculations \citep{seidelmann1992explanatory}.

The analytical models implemented in \texttt{solarsystem} follow the long-established approach of representing planetary motion through time-dependent orbital elements and perturbation terms rather than direct numerical integration. Such methods have been widely used in celestial mechanics because they provide a practical balance between computational efficiency and physical realism \citep{laskar1986secular}.

Time handling within the package relies on astronomical date representations, including Julian dates and fractional day calculations, which provide a continuous temporal framework for orbital computations. Coordinate transformations between spherical and Cartesian systems are employed internally to simplify positional calculations and support conversions between heliocentric, geocentric, ecliptic, and equatorial reference frames \citep{meeus1991astronomical, duffett2017practical}.

Lunar phase and illumination calculations are derived from the relative geometric configuration of the Sun, Earth, and Moon. The illuminated fraction of the lunar disk is estimated from the phase angle between the Sun and Moon as observed from Earth, allowing approximation of the visible illuminated percentage over time \citep{duffett2017practical}.

Rise and set computations are based on horizon-crossing geometry involving the apparent altitude of celestial objects relative to an observer's geographic latitude and longitude. These calculations combine Earth rotation, object declination, and local sidereal geometry to estimate the times at which celestial bodies cross the local horizon.

The orbital calculations implemented in \texttt{solarsystem} are referenced to published sets of orbital elements associated with specific epochs. For the major planets and Pluto, orbital elements referenced to the J2000 epoch (1 January 2000) were adopted. For Ceres, Eris, and Chiron, the implemented elements correspond to more recent reference epochs obtained from available orbital datasets. The use of different reference epochs reflects the availability and quality of published orbital solutions for these bodies and does not affect the general computational methodology. Since the analytical propagation is performed relative to the adopted orbital elements, the resulting calculations remain internally consistent within the intended accuracy range of the package.

To improve positional accuracy beyond simple two-body orbital approximations, \texttt{solarsystem} incorporates several analytical perturbation corrections commonly used in classical positional astronomy \citep{karttunen2007fundamental, montenbruck2013astronomy}. Planetary calculations include perturbative corrections for the mutual gravitational interactions of the outer planets, particularly the dominant Jupiter-Saturn perturbations and additional Uranian terms. Similarly, for lunar calculations, the implemented model includes corrections associated with the Moon's mean elongation, evection, variation, annual equation, and parallactic terms, among other periodic perturbations that account for the gravitational influence of the Earth-Moon-Sun system. The inclusion of these analytical corrections significantly improves agreement with modern ephemerides while preserving the lightweight and dependency-free nature of the computational framework.

Although the package employs simplified analytical models rather than high-order numerical integrations, the implemented methods provide computationally efficient approximations that are sufficiently accurate for educational visualization, observational planning, and exploratory astronomical analysis.

\subsection{Moonrise and Moonset Time Determination}
Moonrise and moonset times are determined using a two-stage iterative estimation procedure. For a given date, an initial approximation is first obtained by evaluating the lunar position at a reference epoch near local noon. The approximate times of moonrise and moonset are then estimated by offsetting this epoch by one-half of the expected lunar transit interval before and after the time of maximum lunar elevation. These preliminary estimates provide a first-order approximation of the horizon-crossing events.

In a second refinement stage, the estimated rise and set times obtained from the initial calculation are used as updated evaluation epochs for recalculating the lunar position and associated horizon-crossing geometry. By recomputing the event times using these improved temporal estimates, the method effectively performs a single iterative correction step that reduces the error introduced by the initial approximation. This approach achieves a favorable balance between computational efficiency and predictive accuracy, avoiding the need for more computationally intensive root-finding algorithms while still providing rise and set times that closely match modern reference ephemerides.

The adopted rise/set determination procedure follows the general principle of successive approximation commonly employed in astronomical event calculations \citep{meeus1991astronomical,montenbruck2013astronomy,seidelmann1992explanatory}. An initial estimate of the event time is first obtained from an approximate lunar configuration and is subsequently refined through recalculation at the estimated epoch. Such iterative approaches are widely used in positional astronomy because the apparent coordinates of celestial bodies vary continuously with time, making a single evaluation insufficient for precise horizon-crossing predictions. By applying a refinement step to the preliminary estimate, the method improves temporal accuracy while maintaining low computational cost, thereby avoiding the need for more complex numerical root-finding procedures.

\subsection{Software Architecture}
\label{ss_repo}

The package is organized into modular components responsible for heliocentric and geocentric positioning, lunar calculations, rise/set prediction, and coordinate transformations. Automated tests are provided to support verification of the implemented algorithms, while accompanying documentation and example notebooks \citep{kluyvertextordfeminine12016jupyter} facilitate reproducibility of the presented workflows. Finally, validation notebooks are included to facilitate reproducibility and to document the accuracy assessments of \texttt{solarsystem} presented in this work.

The core implementation is written entirely in Python and does not require external scientific libraries for its primary functionality. This architecture minimizes software dependencies while maintaining portability across diverse computational environments.

\subsection{Design Philosophy 
and Constraints}

\texttt{Solarsystem} is designed around portability, computational efficiency, and minimal software dependencies. The framework employs analytical methods to compute planetary positions and astronomical events without requiring external ephemerides or scientific libraries. Its modular architecture separates positional astronomy, lunar calculations, rise/set prediction, and coordinate transformations, facilitating maintenance and integration into broader computational workflows. While not intended for high-precision astrometry, the adopted approach provides a practical balance between numerical accuracy, computational cost, and ease of deployment.

The software design also emphasizes reproducibility and transparency, principles that have become increasingly important within computational science \citep{wilson2014best}. By maintaining a compact codebase, minimizing external dependencies, and providing publicly available validation notebooks and test suites, \texttt{solarsystem} seeks to facilitate independent verification of reported results and long-term software sustainability.

\subsection{Ease of Use and Implementation}

\begin{figure}[ht!]
    \centering
    \includegraphics[width=0.95\textwidth,trim=0 0 0 0, clip]{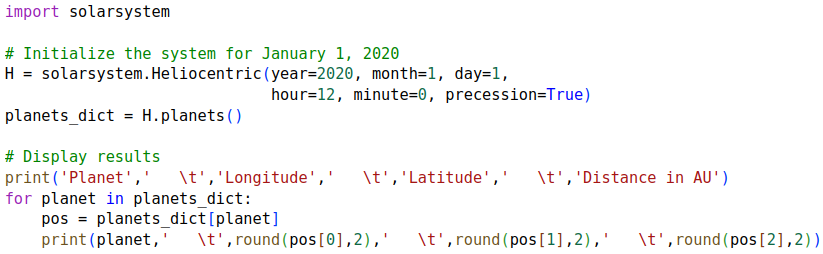}
    \caption{Solarsystem Heliocentric planet position calculation example}
    \label{fig:solarsystem_heliocentric_sample_code}
\end{figure}

To illustrate the practical usage of \texttt{solarsystem}, \autoref{fig:solarsystem_heliocentric_sample_code} presents a complete example for retrieving planetary position data. In this example, a \texttt{Heliocentric} object is instantiated for a specified epoch (1 January 2020, 12:00 UTC), with precession corrections enabled.

A single method call is then used to obtain heliocentric coordinates for all major planets, the dwarf planets Pluto, Ceres, and Eris, and the Centaur Chiron. The resulting output consists of heliocentric longitude, latitude, and radial distance expressed in astronomical units (AU) for each body, as summarized in \autoref{tab:output_results}. The example demonstrates that comprehensive Solar System positional data can be generated with only a small amount of application code, facilitating integration into larger computational workflows and scientific applications.

\begin{table}[h]
\centering
\caption{Output results generated from the implementation example.}
\label{tab:output_results}
\begin{tabular}{lrrr}
\hline
\textbf{Planet} & \textbf{Longitude ($^\circ$)} & \textbf{Latitude ($^\circ$)} & \textbf{Distance (AU)} \\ \hline
Mercury & 263.55 & -4.06 & 0.47 \\
Venus   & 4.95   & -3.22 & 0.73 \\
Earth   & 100.25 & 0.00  & 0.98 \\
Mars    & 214.10 & 0.49  & 1.59 \\
Jupiter & 275.83 & 0.10  & 5.23 \\
Saturn  & 292.23 & 0.05  & 10.05 \\
Uranus  & 35.07  & -0.48 & 19.81 \\
Neptune & 347.74 & -1.04 & 29.91 \\
Pluto   & 292.47 & -0.67 & 33.88 \\
Ceres   & 290.44 & -5.40  & 2.92 \\
Chiron  & 3.86   & 2.94  & 18.81 \\
Eris    & 23.08  & -11.74 & 96.00 \\ \hline
\end{tabular}
\end{table}

\section{Validation}
\label{validation}

\subsection{Positional Accuracy Assessment}
\label{positional_accuracy}

The numerical accuracy of \texttt{solarsystem} was evaluated through a series of validation experiments against the JPL DE440 planetary ephemerides \citep{park2021jpl,folkner2014planets}, accessed via the \texttt{Skyfield} framework. The comparison included all eight major planets and Pluto and considered heliocentric longitude, heliocentric latitude, and heliocentric distance expressed in astronomical units (AU). The results demonstrate good agreement with the reference ephemerides across the examined epochs, indicating that the adopted analytical approach provides reliable planetary position estimates while maintaining a lightweight computational design.

The DE440 ephemerides were selected as the reference dataset because they represent the current high-precision planetary and lunar ephemerides developed by JPL and are widely adopted throughout the astronomical community.

Two independent experiments were conducted. The first one evaluated planetary positions for every day of the year 2025 at 12:00 UTC, while the second examined long-term behavior over the interval 1950-2050 using one sample per year on January 1\textsuperscript{st}. For each epoch, positional outputs generated by \texttt{solarsystem} were compared directly against the corresponding DE440 reference values.

The resulting average deviations across all nine planetary bodies are summarized in ~\autoref{table:validation_results}. As shown in the table, the observed positional differences remained consistently small for both experimental intervals. Mean absolute heliocentric longitude deviations were found to remain on the order of less than one arcminute, while heliocentric latitude differences were even smaller. Radial distance deviations remained modest considering the simplified and lightweight computational framework employed by the package.

\begin{table}[h]
\caption{Validation results - Mean Absolute Error, averaged over all planets}
\label{table:validation_results}
\centering
\begin{tabular}{l ccc} 
 \hline
 Measurement & Mean Longitude diff. & Mean Latitude diff. & Mean Distance diff. \\ 
 \hline
Year 2025 & 0° 0.48' & 0° 0.13' & 0.06425205 AU \\ 
1950-2050 & 0° 0.44' & 0° 0.16' & 0.05035441 AU \\ 

\end{tabular}
\end{table}

A more detailed breakdown of the validation results is presented in \autoref{table:validation_results_perplanet}, which reports the mean, maximum, and standard deviation of the longitude, latitude, and heliocentric distance differences for each planet. The largest longitude deviations are observed for Saturn and Uranus, while Saturn and Jupiter exhibit the largest latitude discrepancies. Saturn also shows the greatest deviation in heliocentric distance. These trends are consistent with the stronger mutual gravitational perturbations experienced by the outer planets, particularly within the Jupiter-Saturn-Uranus system. Despite exhibiting the largest deviations among the examined bodies, the observed errors remain small in absolute terms, indicating that the implemented analytical models provide reliable positional estimates for a wide range of Solar System calculations and exploratory celestial mechanics applications.

\begin{table}[h]
\caption{Validation results - Mean Absolute Error per planet for the period 1950-2050 }
\label{table:validation_results_perplanet}
\centering
\begin{tabular}{l |ccc|ccc|ccc} 
 
 \toprule
 Planet
& \multicolumn{3}{c|}{\textbf{Longitude difference}}
& \multicolumn{3}{c|}{\textbf{Latitude difference}}
& \multicolumn{3}{c}{\textbf{Distance difference}}\\

  & Mean & Max & StD & Mean & Max & StD & Mean & Max & StD\\ 
 \midrule
 Mercury & 0.32' & 0.82' & 0.13' & 0.13' & 0.38' & 0.10' & 0.0015 & 0.0052 & 0.0011 \\
Venus & 0.31' & 0.67' & 0.20' & 0.13' & 0.39' & 0.10' & 0.0020 & 0.0045 & 0.0011 \\
Earth & 0.40' & 0.68' & 0.13' & 0.19' & 0.38' & 0.11' & 0.0015 & 0.0047 & 0.0011 \\
Mars & 0.47' & 1.51' & 0.33' & 0.12' & 0.34' & 0.09' & 0.0053 & 0.0146 & 0.0037 \\
Jupiter & 0.37' & 1.17' & 0.23' & 0.20' & 0.73' & 0.16' & 0.0375 & 0.1319 & 0.0306 \\
Saturn & 0.62' & 1.68' & 0.45' & 0.24' & 0.66' & 0.17' & 0.1750 & 0.4387 & 0.1102 \\
Uranus & 0.68' & 1.64' & 0.54' & 0.12' & 0.33' & 0.10' & 0.0502 & 0.1091 & 0.0271 \\
Neptune & 0.44' & 1.22' & 0.30' & 0.13' & 0.35' & 0.09' & 0.0489 & 0.0768 & 0.0144 \\
Pluto & 0.37' & 0.88' & 0.25' & 0.19' & 0.49' & 0.12' & 0.1314 & 0.2547 & 0.0635 \\

 \bottomrule
 \multicolumn{10}{l}{\footnotesize{ Longitude and Latitude differences in \textbf{arcminutes}. Distance differences in AU. }} \\
\end{tabular}
\end{table}

Additional validation experiments were performed to assess the accuracy of the implemented lunar position calculations. The same two temporal intervals used for the planetary validation (daily evaluations throughout 2025 and annual evaluations spanning 1950-2050) were also applied to the Moon. As shown in \autoref{table:validation_results_moon}, the mean difference relative to the reference ephemerides remains below 1.5 arcminutes in longitude and below 1 arcminute in latitude. Although the lunar calculations exhibit somewhat larger positional deviations than those observed for most planets, this behavior is expected given the greater complexity of lunar motion and the stronger influence of periodic perturbations within the Earth-Moon-Sun system. Nevertheless, the overall agreement with the reference ephemerides remains strong. Furthermore, the discrepancies in the Earth-Moon distance, expressed in astronomical units (AU), are considerably smaller than those reported for the planets, reflecting the substantially shorter Earth-Moon distance compared with typical heliocentric planetary distances.

\begin{table}[h]
\caption{Validation results for Moon position - Mean Absolute Error}
\label{table:validation_results_moon}
\centering
\begin{tabular}{l |ccc|ccc|ccc} 
 
 \toprule
 Experiment
& \multicolumn{3}{c|}{\textbf{Longitude diff.}}
& \multicolumn{3}{c|}{\textbf{Latitude diff.}}
& \multicolumn{3}{c}{\textbf{Distance diff.}}\\

  & Mean & Max & StD & Mean & Max & StD & Mean & Max & StD\\ 
 \midrule
 2025 & 1.40' & 4.34' & 0.93' & 0.93' & 2.90' & 0.62' & 2.046 & 4.153 & 1.151 \\
1950-2050 & 1.40' & 4.32' & 1.09' & 0.87' & 2.78' & 0.59' & 1.507 & 2.985 & 0.798 \\

 \bottomrule
 \multicolumn{10}{l}{\footnotesize{ Longitude and Latitude differences in \textbf{arcminutes}. Distance differentials of \num{1e-5}  astronomical units (AU). }} \\
\end{tabular}
\end{table}

The consistency of the results over both short-term and century-scale intervals suggests that the implemented analytical models provide stable and numerically reliable approximations for the intended application domain of the library. While the package does not attempt to reproduce the sub-arcsecond precision of professional numerical ephemerides, the observed agreement with DE440 demonstrates that \texttt{solarsystem} achieves a practical balance between computational simplicity, portability, and astronomical realism.

Part of the remaining systematic positional deviation is likely attributable to the simplified orbital models and the omission of high-order coordinate corrections such as nutation, relativistic effects, and detailed perturbation modeling. Nevertheless, the achieved accuracy is considered sufficient for educational visualization, exploratory numerical analysis, observational planning, and interactive Solar System simulations.

\subsection{Solar-Lunar Event Prediction Accuracy}
\label{positional_accuracy}

Validation experiments were performed for lunar rise and set predictions, as well as lunar illumination estimates, using the JPL DE440 ephemerides as the reference solution. The same temporal intervals employed for the planetary validation were considered, namely daily evaluations throughout 2025 and annual evaluations spanning the period 1950-2050. Unlike the planetary position comparisons, the rise/set analysis is reported using percentile statistics rather than mean and maximum errors. This choice was motivated by the periodic nature of rise and set events: in rare cases where an event occurs close to midnight, two otherwise equivalent predictions may be assigned to adjacent calendar days, producing apparent timing differences approaching 24 hours. Such occurrences represent a date-boundary ambiguity rather than a true prediction error and can disproportionately influence conventional summary statistics. Percentile-based measures therefore provide a more representative assessment of the typical agreement between the predicted and reference event times.

\begin{table}[h]
\caption{Validation results for Moon rise/set and illumination - Mean Absolute Error}
\label{table:validation_results_moon_risesetillum}
\centering
\begin{tabular}{l |ccc|ccc|ccc} 
 
 \toprule
 Experiment
& \multicolumn{3}{c|}{\textbf{Moonrise}}
& \multicolumn{3}{c|}{\textbf{Moonset}}
& \multicolumn{3}{c}{\textbf{Illumination (\%)}}\\

  & 25\% & 50\% & 75\% & 25\% & 50\% & 75\% & Mean & Max & StD\\ 
 \midrule
 2025 & 1.50 & 2.75 & 3.52 & 1.43 & 2.70 & 3.58 & 0.18 & 0.60 & 0.14 \\
1950-2050 & 1.42 & 2.33 & 3.22 & 1.15 & 2.32 & 3.17 & 0.26 & 0.60 & 0.19 \\

 \bottomrule
 \multicolumn{10}{l}{\footnotesize{ Error at different percentiles for Moonrise and Moonset expressed in minutes.}} \\
\end{tabular}
\end{table}

As summarized in \autoref{table:validation_results_moon_risesetillum}, the agreement between \texttt{solarsystem} and the reference ephemerides is strong, with median timing differences of approximately 2-3 minutes for both moonrise and moonset events and 75th-percentile deviations generally below 4 minutes. Similar results were obtained for both validation periods, indicating consistent performance across short- and long-term timescales. Lunar illumination calculations exhibited very small discrepancies relative to the reference ephemerides, with mean differences of approximately 0.2\%, maximum deviations below 0.6\%, and standard deviations below 0.2\%. These results indicate that the implemented lunar phase model reproduces the illuminated fraction of the Moon with a high degree of accuracy.

The remaining discrepancies relative to the DE440 ephemerides arise primarily from the analytical approximations employed by the implemented orbital models. Additional contributions may originate from neglected higher-order perturbations, simplified lunar theories, and differences in reference frame conventions. Such effects are expected given the lightweight design philosophy of the framework and remain small relative to the intended application domain.

Additional validation was performed for sunrise and sunset predictions using the same temporal intervals considered throughout this study, namely daily evaluations during 2025 and annual evaluations spanning 1950-2050. Unlike lunar rise and set events, sunrise and sunset occur sufficiently far from midnight to avoid date-boundary ambiguities, allowing conventional summary statistics to provide a representative assessment of prediction accuracy. For the 2025 experiment, the mean, maximum, and standard deviation of the timing differences relative to the reference ephemerides were 0.27, 0.52, and 0.13 minutes, respectively. For the 1950-2050 experiment, the corresponding values were 0.06, 0.25, and 0.07 minutes. These results demonstrate excellent agreement with the reference ephemerides, with typical discrepancies remaining well below one minute across both validation periods, indicating that the implemented solar rise/set algorithms provide highly accurate predictions while maintaining the lightweight computational design of the package.

All topographic validation experiments were performed using the geographic coordinates of Athens, Greece, with Universal Time (UT) as the temporal reference. Additional experiments conducted using the coordinates of Greenwich, United Kingdom, produced comparable results, suggesting that the observed deviations primarily reflect differences between the computational models rather than location-specific effects. Overall, the results demonstrate that the implemented lunar algorithms provide reliable predictions for observational planning and related astronomical applications while maintaining the lightweight computational design of the package.

An important aspect of the present work is the explicit quantitative validation of the implemented algorithms against the JPL DE440 ephemeris. While many astronomical software packages provide positional calculations, published assessments of their deviations from high-precision reference ephemerides are often unavailable or limited in scope. By reporting systematic comparisons across multiple bodies and extended temporal intervals, this study provides users with a clear understanding of the expected accuracy and limitations of the package. The results demonstrate that the implemented analytical methods achieve good agreement with modern reference ephemerides while maintaining a lightweight computational framework.

\subsection{Computational Performance}

To assess computational efficiency, a benchmark was performed by calculating heliocentric positions for all major bodies currently supported by \texttt{solarsystem}, including the eight planets, the dwarf planets Pluto, Ceres, and Eris, and the Centaur object Chiron. Calculations were performed for every day over a 100-year interval spanning 1950-2050, corresponding to approximately 36,525 distinct epochs and a total execution time of 1.4 seconds on a standard Google Colab environment. This corresponds to an average computational cost of approximately ($3.9 \times 10^{-5}$) per epoch for the complete set of supported bodies. The results demonstrate that the analytical approach adopted by \texttt{solarsystem} enables rapid large-scale temporal analyses while maintaining a lightweight implementation free from external ephemeris dependencies.

\section{Impact of Precession Modeling on Planetary Longitude Accuracy}

\begin{figure}[ht!]
    \centering
    \includegraphics[width=0.75\textwidth,trim=0 0 0 0, clip]{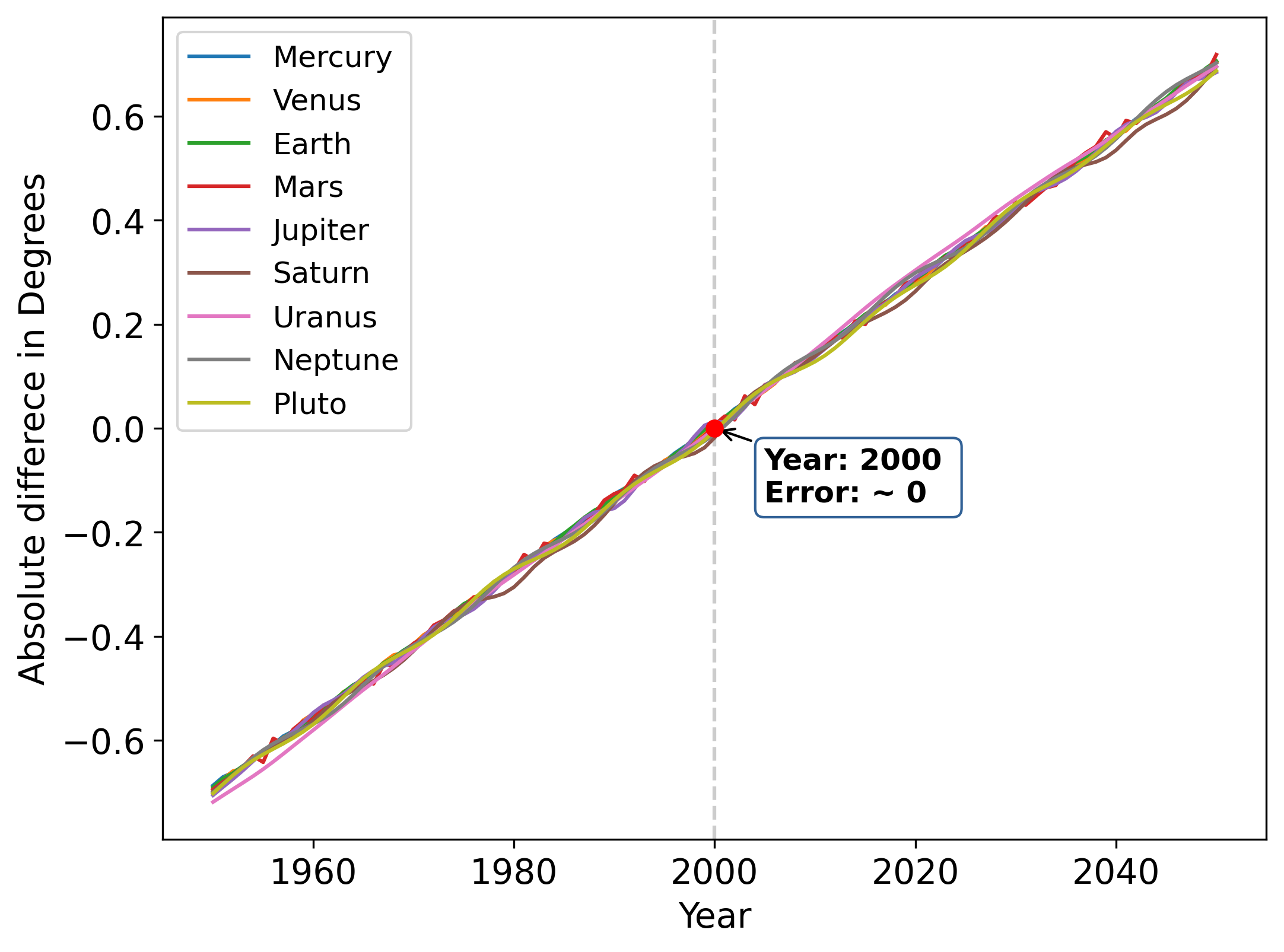}
    \caption{Planet Heliocentric position difference without precession correction}
    \label{fig:planet_logitude_precession_deviation}
\end{figure}

Additional validation experiments were conducted using positional calculations in which the optional longitude precession correction was disabled. The resulting longitude deviations relative to the reference ephemeris are illustrated in \autoref{fig:planet_logitude_precession_deviation}, which presents the heliocentric longitude differences for all examined planetary bodies over the investigated temporal interval.

As shown in the figure, the longitude deviations exhibit a clear systematic temporal trend consistent with the expected effects of the precession of the equinoxes. Near the J2000 reference epoch, the observed longitude differences remain close to zero for all planets, indicating strong agreement between the \texttt{solarsystem} computational framework and the reference ephemeris when both coordinate systems are effectively aligned to the same epoch. Moving away from the year 2000, however, the positional differences gradually increase in magnitude.

For epochs near 1950, the longitude deviations become increasingly negative, corresponding to cases where the reference ephemeris produce slightly larger longitude values than those computed by \texttt{solarsystem}. Conversely, toward epochs near 2050, the deviations become increasingly positive, indicating that \texttt{solarsystem} yields larger longitude values relative to the ephemeris. The overall magnitude of the observed discrepancies remains relatively small, typically remaining below one degree with an average of approximately 0.35 degrees for the period between 1950 and 2050.

The observed behavior strongly suggests that the dominant systematic component of the positional discrepancy originates from the omission of epoch-dependent precession corrections rather than from instability in the underlying orbital calculations themselves. This trend is physically consistent with the gradual rotational drift of the terrestrial reference frame caused by Earth's axial precession. The results therefore support the interpretation that the implemented orbital models remain internally stable while highlighting the importance of coordinate-frame evolution in long-term astronomical comparisons.

\section{Application Examples}

\begin{figure}[ht!]
    \centering
    \includegraphics[width=0.85\textwidth,trim=0 0 0 0, clip]{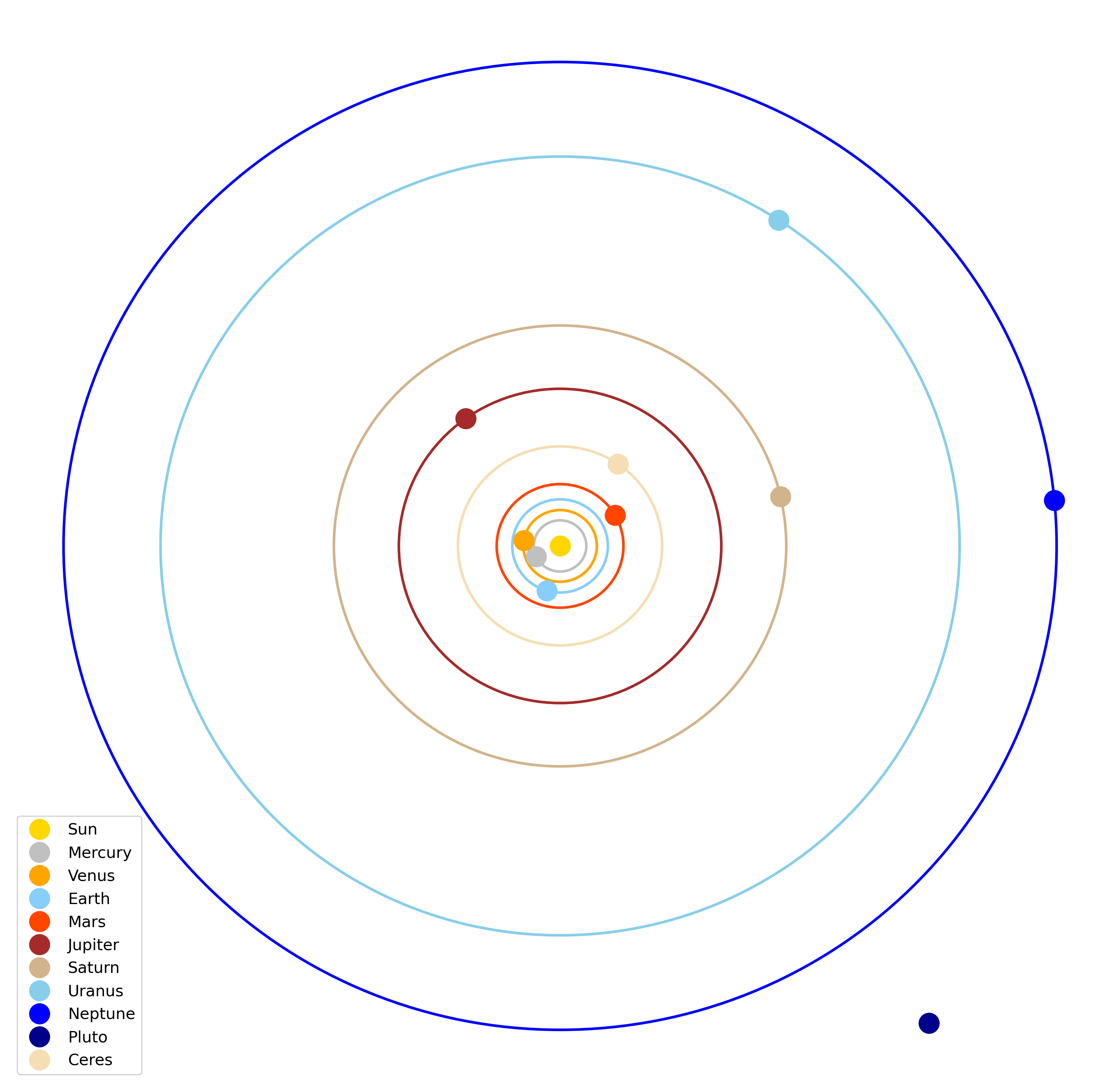}
    \caption{Planets position around the Sun on 11/06/2026}
    \label{fig:solar_system_live}
\end{figure}

Beyond direct positional calculations, \texttt{solarsystem} enables the creation of visualization and long-term astronomical analysis workflows. The package supports both real-time and time-parameterized Solar System calculations, allowing users to compute heliocentric or geocentric planetary positions for arbitrary epochs in the past, present, or future. This functionality enables the generation of live Solar System representations, orbital animations, and retrospective or predictive studies of planetary configurations. An example visualization is shown in \autoref{fig:solar_system_live}. Such capabilities are relevant to long-term planetary monitoring, astronomical visualization, observational planning, and exploratory studies of Solar System dynamics.

\begin{figure}[ht!]
    \centering
    \includegraphics[width=0.99\textwidth,trim=0 0 0 0, clip]{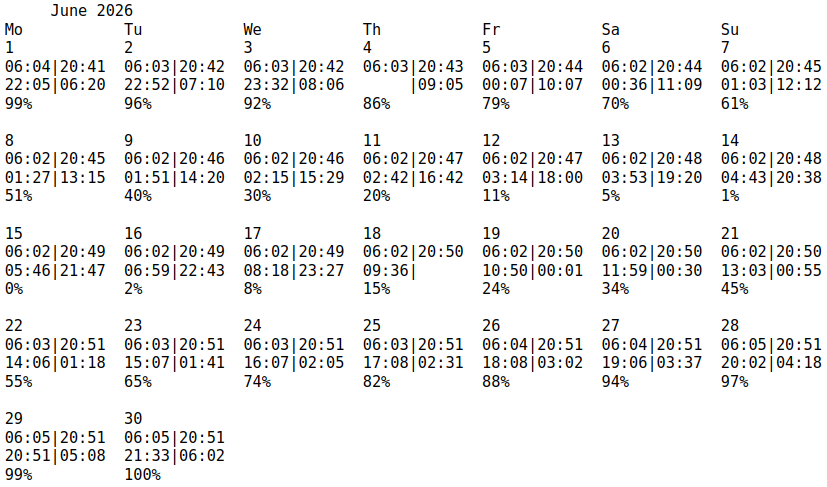}
    \caption{June 2026 calendar for Athens, Greece. Sunrise/set, Moonrise/set and lunar illumination}
    \label{fig:riseset_calendar}
\end{figure}

In addition to planetary visualization, the package can generate extended astronomical calendars containing sunrise, sunset, moonrise, moonset, and lunar illumination information for user-defined geographic locations. The generated calendars provide compact observational datasets that can support site-specific observing schedules and temporal analyses of solar and lunar phenomena. An example rise/set and lunar illumination calendar generated using the package is presented in \autoref{fig:riseset_calendar}.

Both workflows are demonstrated through dedicated Jupyter notebooks included in the repository's \texttt{examples} directory. The \href{https://github.com/IoannisNasios/solarsystem/blob/master/examples/Solar_System_Live.ipynb}{\textit{Solar System Live}} notebook illustrates dynamic planetary visualization, while the \href{https://github.com/IoannisNasios/solarsystem/blob/master/examples/RiseSet_Calendar.ipynb}{\textit{RiseSet Calendar}} notebook demonstrates automated generation of astronomical observation calendars and long-term time-series analyses involving solar and lunar phenomena.

\section{Limitations and Future Work}
\label{limitations}

The package employs analytical approximations intended to balance computational efficiency and numerical accuracy. Consequently, it should not be regarded as a replacement for high-precision numerical ephemerides in applications requiring sub-arcsecond astrometry, spacecraft navigation, or mission design.

As a consequence of its lightweight architecture, the positional accuracy of the package is generally lower than that achieved by modern JPL Development Ephemerides and related professional astronomical frameworks. Simplifications in orbital modeling, lunar phase estimation, and rise/set calculations may introduce measurable deviations, particularly over long temporal intervals or in applications requiring highly precise topocentric corrections.

The current implementation also omits several advanced physical effects commonly incorporated into professional astronomical software, including other relativistic corrections, detailed perturbation modeling, atmospheric refraction refinements, and high-order numerical integrations. In addition, support for minor Solar System bodies remains limited compared to specialized ephemeris systems that include extensive asteroid and comet catalogs.

These limitations are consistent with the primary design goals of the project, which emphasize portability, dependency-free, ease of installation, and lightweight computation rather than maximal astrometric precision. Within this intended application scope, the package nevertheless provides sufficiently accurate results for visualization, educational demonstrations, observational planning, and exploratory numerical analysis.

\section{Conclusion}
\label{Conclusion}

This work introduced \texttt{solarsystem}, a lightweight and dependency-free Python package developed for planetary positions and solar-lunar event calculations, and astronomical coordinate transformations. The project emphasizes simplicity, portability, and ease of use, enabling meaningful astronomical computations without requiring external ephemeris datasets, large scientific infrastructures, or complex installation procedures. The pure Python implementation allows straightforward integration into educational environments, lightweight applications, and exploratory scientific workflows.

The package provides functionality for heliocentric and geocentric planetary positioning, lunar illuminated fraction estimation, rise and set calculations, and coordinate system conversions. These capabilities support a broad range of applications including educational demonstrations, observational planning, visualizations, and exploratory numerical analysis. In addition, \texttt{solarsystem} enables real-time and time-parameterized Solar System visualization together with automated generation of long-term astronomical calendars containing sunrise, sunset, moonrise, moonset, and lunar illumination information.

Validation experiments performed against JPL DE440 reference ephemeris demonstrated that the implemented computational methods achieve satisfactory agreement for the intended application domain. Mean planetary positional deviations remained below half an arcminute for the examined validation intervals, while yearly average moonrise and moonset timing differences were found to be around 2.5 minutes after temporal reference alignment. Lunar illumination calculations exhibited average deviations of roughly 0.2\%, indicating that the simplified computational framework remains sufficiently reliable for visualization, educational, and observational use cases.

An important feature of the package is the optional inclusion of precession corrections associated with the Earth's axial motion. While precession is enabled by default in order to maintain consistency with modern astronomical reference frames and ephemerides, users may explicitly disable this correction when working within a fixed coordinate framework. This flexibility is particularly valuable for comparative orbital studies, long-term visualization workflows, and educational demonstrations where maintaining stable relative coordinates across epochs may be preferable to applying continuously evolving terrestrial reference-frame corrections.

While not intended to replace high-precision ephemeris systems in mission-critical astrodynamics applications, \texttt{solarsystem} demonstrates that lightweight analytical methods can provide scientifically useful accuracy while maintaining computational efficiency, portability, and ease of deployment. By combining validated planetary and solar-lunar calculations within a dependency-free Python package, the project contributes a reproducible and accessible tool for computational astronomy. Future development may focus on incorporating additional minor bodies, refining precession and perturbation models, and further improving numerical accuracy through comparison with modern reference ephemerides.

\section*{Declaration of Competing Interest}
The author declares that has no known competing financial interests or personal relationships that could have appeared to influence the work reported in this paper.

\section*{Code Availability}
\label{sec:Code availability} 

Version 0.1.8 of the \texttt{solarsystem} package was used throughout this study and is publicly available via PyPI and the project's GitHub repository. Example notebooks demonstrating the applications presented in this paper, and validation scripts and notebooks used to reproduce the ephemeris comparison results are available at the project's GitHub repository \url{https://github.com/IoannisNasios/solarsystem}. In this repository, source code and documentation are also openly available to support reproducibility and future development.

\bibliographystyle{cas-model2-names}

\bibliography{refs}

@article{astropy2013,
  author = {{Astropy Collaboration}},
  title = {Astropy: A community Python package for astronomy},
  journal = {Astronomy \& Astrophysics},
  volume = {558},
  pages = {A33},
  year = {2013}
}

@article{poliastro,
  author = {Juan Luis Cano Rodríguez and others},
  title = {poliastro: Python library for interactive astrodynamics},
  journal = {Journal of Open Source Software},
  volume = {5},
  number = {49},
  pages = {2187},
  year = {2020}
}

@misc{spiceypy,
  author = {Andrew Annex and others},
  title = {SpiceyPy: a Pythonic Wrapper for the SPICE Toolkit},
  year = {2020},
  publisher = {Zenodo},
  doi = {10.5281/zenodo.3826081}
}

@misc{skyfield,
  author = {Brandon Rhodes},
  title = {Skyfield: High precision research-grade positions for planets and Earth satellites generator},
  year = {2019},
  howpublished = {\url{https://rhodesmill.org/skyfield/}}
}

@article{folkner2014planets,
  author = {William M. Folkner and James G. Williams and Dale H. Boggs and Ryan S. Park and Petr Kuchynka},
  title = {The Planetary and Lunar Ephemerides DE430 and DE431},
  journal = {Interplanetary Network Progress Report},
  volume = {42-196},
  pages = {1--81},
  year = {2014}
}

@article{kuang2023simulations, title={Simulations of triple microlensing events I: detectability of a scaled Sun--Jupiter--Saturn system}, author={Kuang, Renkun and Zang, Weicheng and Mao, Shude and Zhang, Jiyuan and Jiang, Haochang}, journal={Monthly Notices of the Royal Astronomical Society}, volume={520}, number={3}, pages={4540--4553}, year={2023}, publisher={Oxford University Press} }

@article{park2021jpl,
  title={The JPL planetary and lunar ephemerides DE440 and DE441},
  author={Park, Ryan S and Folkner, William M and Williams, James G and Boggs, Dale H},
  journal={The Astronomical Journal},
  volume={161},
  number={3},
  pages={105},
  year={2021},
  publisher={The American Astronomical Society}
}

@book{meeus1991astronomical,
  title={Astronomical algorithms},
  author={Meeus, Jean H},
  year={1991},
  publisher={Willmann-Bell, Incorporated}
}

@article{simon1994numerical,
  title={Numerical expressions for precession formulae and mean elements for the Moon and the planets},
  author={Simon, Jean-Louis and Bretagnon, P and Chapront, J and Chapront-Touze, M and Francou, G and Laskar, J},
  journal={Astronomy and Astrophysics (ISSN 0004-6361), vol. 282, no. 2, p. 663-683},
  volume={282},
  pages={663--683},
  year={1994}
}

@book{seidelmann1992explanatory,
  title={Explanatory supplement to the astronomical almanac},
  author={Seidelmann, P Kenneth},
  year={1992},
  publisher={University Science Books}
}

@book{duffett2017practical,
  title={Practical astronomy with your calculator or spreadsheet},
  author={Duffett-Smith, Peter and Zwart, Jonathan},
  year={2017},
  publisher={Cambridge University Press}
}

@book{karttunen2007fundamental,
  title={Fundamental astronomy},
  author={Karttunen, Hannu and Kr{\"o}ger, Pekka and Oja, Heikki and Poutanen, Markku and Donner, Karl Johan},
  year={2007},
  publisher={Springer}
}

@book{montenbruck2013astronomy,
  title={Astronomy on the personal computer},
  author={Montenbruck, Oliver and Pfleger, Thomas},
  year={2013},
  publisher={Springer}
}

@article{berger1978long,
  title={Long-term variations of daily insolation and Quaternary climatic changes},
  author={Berger, Andr{\'e}L},
  journal={Journal of Atmospheric Sciences},
  volume={35},
  number={12},
  pages={2362--2367},
  year={1978}
}

@article{rowe2015galsim,
  title={GALSIM: The modular galaxy image simulation toolkit},
  author={Rowe, Barnaby TP and Jarvis, Mike and Mandelbaum, Rachel and Bernstein, Gary M and Bosch, James and Simet, Melanie and Meyers, Joshua E and Kacprzak, Tomasz and Nakajima, Reiko and Zuntz, Joe and others},
  journal={Astronomy and Computing},
  volume={10},
  pages={121--150},
  year={2015},
  publisher={Elsevier}
}

@article{akeret2015hope,
  title={HOPE: A Python just-in-time compiler for astrophysical computations},
  author={Akeret, Jo{\"e}l and Gamper, Lukas and Amara, Adam and Refregier, Alexandre},
  journal={Astronomy and Computing},
  volume={10},
  pages={1--8},
  year={2015},
  publisher={Elsevier}
}

@inproceedings{kluyvertextordfeminine12016jupyter,
  title={Jupyter Notebooks-a publishing format for reproducible computational workflows},
  author={KLUYVER{\textordfeminine}$^1$, Thomas and RAGAN-KELLEYb$^1$, Benjamin and P{\'e}rez, Fernando and Granger, Brian and Bussonnier, Matthias and Frederic, Jonathan and Kelley, Kyle},
  booktitle={Positioning and power in academic publishing: players, agents and agendas: proceedings of the 20th International Conference on Electronic Publishing},
  pages={87},
  year={2016},
  organization={IOS press}
}

@article{laskar1986secular,
  title={Secular terms of classical planetary theories using the results of general theory},
  author={Laskar, Jacques},
  journal={Astronomy and astrophysics},
  volume={157},
  number={1},
  pages={59--70},
  year={1986}
}

@article{wilson2014best,
  title={Best Practices for Scientific Computing},
  author={Wilson, Greg and others},
  journal={PLoS Biology},
  volume={12},
  number={1},
  pages={e1001745},
  year={2014}
}

\end{document}